\documentclass{article}

\usepackage{amsmath,amssymb}
\usepackage{graphicx}
\usepackage{color}
\usepackage[all]{xy}

\newtheorem{theorem}{Theorem}
\newtheorem{example}{Example}
\newtheorem{lemma}{Lemma}
\newtheorem{corollary}{Corollary}

\newtheorem{proposition}{Proposition}
\newtheorem{remark}{Remark}

\newcommand{\done}{\hfill $\Box$ }

\newcommand{\ls}[1]
    {\dimen0=\fontdimen6\the\font\lineskip=#1\dimen0
     \advance\lineskip.5\fontdimen5\the\font
     \advance\lineskip-\dimen0
     \lineskiplimit=0.9\lineskip
     \baselineskip=\lineskip
     \advance\baselineskip\dimen0
     \normallineskip\lineskip\normallineskiplimit\lineskiplimit
     \normalbaselineskip\baselineskip
     \ignorespaces}

\begin{document}

\bibliographystyle{abbrv}

\title{New Classes of Ternary Bent Functions from the Coulter-Matthews Bent Functions}

\author{$^1$Honggang Hu, $^1$Xiaolong Yang, and $^2$Shaohua Tang\\
\and
$^1$School of Information Science and Technology\\
University of Science and Technology of China\\
Hefei, China, 230027\\
Email. hghu2005@ustc.edu.cn, yxl@mail.ustc.edu.cn\\
\and
$^2$School of Computer Science and Engineering\\
South China University of Technology\\
Guangzhou, China, 510006\\
Email. csshtang@scut.edu.cn}

\date{}
 \maketitle

\thispagestyle{plain}
\setcounter{page}{1}

\begin{abstract}
It has been an active research issue for many years to construct new bent functions.
For $k$ odd with $\gcd(n, k)=1$, and $a\in\mathbb{F}_{3^n}^{*}$, the function $f(x)=Tr(ax^{\frac{3^k+1}{2}})$ is weakly regular
bent over $\mathbb{F}_{3^n}$, where $Tr(\cdot):\mathbb{F}_{3^n}\rightarrow\mathbb{F}_3$ is the trace function. This is the well-known Coulter-Matthews bent function. In this paper, we determine the dual function of $f(x)$ completely. As a consequence, we find many classes of ternary bent functions not reported in the literature previously. Such bent functions are not quadratic if $k>1$, and have $\left(\left(\frac{1+\sqrt{5}}{2}\right)^{w+1}-\right.$ $\left.\left(\frac{1-\sqrt{5}}{2}\right)^{w+1}\right)/\sqrt{5}$ or $\left(\left(\frac{1+\sqrt{5}}{2}\right)^{n-w+1}-\right.$ $\left.\left(\frac{1-\sqrt{5}}{2}\right)^{n-w+1}\right)/\sqrt{5}$ trace terms, where $0<w<n$ and $wk\equiv 1\ (\bmod\;n)$.
Among them, five special cases are especially interesting: for the case of $k=(n+1)/2$, the number of trace terms is $\left(\left(\frac{1+\sqrt{5}}{2}\right)^{n-1}-\right.$ $\left.\left(\frac{1-\sqrt{5}}{2}\right)^{n-1}\right)/\sqrt{5}$;
for the case of $k=n-1$, the number of trace terms is $\left(\left(\frac{1+\sqrt{5}}{2}\right)^n-\right.$ $\left.\left(\frac{1-\sqrt{5}}{2}\right)^n\right)/\sqrt{5}$;
for the case of $k=(n-1)/2$, the number of trace terms is $\left(\left(\frac{1+\sqrt{5}}{2}\right)^{n-1}-\right.$ $\left.\left(\frac{1-\sqrt{5}}{2}\right)^{n-1}\right)/\sqrt{5}$;
for the case of $(n, k)=(5t+4, 4t+3)$ or $(5t+1, 4t+1)$ with $t\geq 1$, the number of trace terms is 8;
and for the case of $(n, k)=(7t+6, 6t+5)$ or $(7t+1, 6t+1)$ with $t\geq 1$, the number of trace terms is 21.
As a byproduct, we find new classes of ternary bent functions with only 8 or 21 trace terms.
\end{abstract}


\ls{1.5}
\section{Introduction}\label{sec_intro}

Boolean bent functions have the maximum nonlinearity, i.e., the maximum Hamming distance to the
set of all affine functions. Since Dillon \cite{Dillon72} and Rothaus \cite{Rothaus76} introduced Boolean bent functions firstly, these functions have found many applications in cryptography, coding theory and communications \cite{GG05,HK98}. The reader is referred to \cite{Mesnager16} for more history on Boolean bent functions. Bent functions have been an active research issue for around 40 years. In 1985, the concept of Boolean bent functions was generalized to the case of functions over integer residue rings by Kumar {\em et al.} \cite{KuScWe85}. Generalized bent functions
are naturally much more complicated than Boolean bent functions. There are some known constructions over finite fields \cite{CM97-1,HeHoKhWaXi09,HeKh06_1,HeKh07,HeKh10,Ho08,HZS17,KJNH03,KM91,LK92}.

Let $p$ be a prime number and $n\geq 1$. Let $\mathbb{F}_{p^n}$ be the finite field with $p^n$ elements. Suppose that $f(x)$ is a function from $\mathbb{F}_{p^n}$ to $\mathbb{F}_p$. Then the Walsh transform of $f(x)$ is defined by
$$\widehat{f}(\lambda)=\sum_{x\in \mathbb{F}_{p^n}}\omega_p^{f(x)-Tr(\lambda x)},$$
where $Tr(\cdot)$ is the trace function from $\mathbb{F}_{p^n}$ to $\mathbb{F}_p$, $\lambda\in \mathbb{F}_{p^n}$, and $\omega_p=e^{2\pi i/p}$. The inverse Walsh transform of $f(x)$ is defined by
$$\omega_p^{f(x)}=\frac{1}{p^n}\sum_{\lambda\in \mathbb{F}_{p^n}}\widehat{f}(\lambda)\omega_p^{Tr(\lambda x)}.$$
If $|\widehat{f}(\lambda)|^2=p^n$ for any $\lambda\in \mathbb{F}_{p^n}$, then $f(x)$ is called a generalized bent function over $\mathbb{F}_{p^n}$. In this case,
we also call $f(x)$ a $p$-ary bent function. If there exist a function $g(x)$ from $\mathbb{F}_{p^n}$ to $\mathbb{F}_p$ and a complex number $u$ with $|u|=1$ such that
$$\widehat{f}(\lambda)=up^{\frac{n}{2}}\omega_p^{g(\lambda)}$$
for any $\lambda\in \mathbb{F}_{p^n}$, then $f(x)$ is weakly regular bent, and $g(x)$ is the dual function of $f(x)$. In this case, $g(x)$ is also weakly regular bent. Actually, there are only four cases for the constant $u$: $\pm1, \pm i$ \cite{HeKh06_1}. In particular, if $u=1$, i.e.,
$$\widehat{f}(\lambda)=p^{\frac{n}{2}}\omega_p^{g(\lambda)}$$
for any $\lambda\in \mathbb{F}_{p^n}$, then $f(x)$ is regular bent. In this case, $g(x)$ is also regular bent.

For the case of ternary bent functions, there are some known constructions \cite{CM97,CM97-1,HeKh06_1,HZS17,KJNH03,KM91,LK92}. The Coulter-Matthews bent functions are especially interesting. Let $d=(3^k+1)/2$ with $k$ odd and $\gcd(n, k)=1$. In 1997, for any $a\in \mathbb{F}_{3^n}^{*}$, Coulter and Matthews showed that $Tr(ax^d)$ is bent over $\mathbb{F}_{3^n}$ \cite{CM97,CM97-1}. This is the well-known Coulter-Matthews bent function. In 2006, Helleseth and Kholosha proposed two conjectures \cite{HeKh06_1}: 1) the Coulter-Matthews bent functions are weakly regular bent;
2) another construction of monomial bent functions over $\mathbb{F}_{3^n}$. In 2007, Feng and Luo investigated one class of exponential sums
associated with the Coulter-Matthews bent functions, and determined the value distributions \cite{FL07}.
For the first conjecture proposed by Helleseth and Kholosha, Hou proved two special cases in 2008 \cite{Ho08}, and
Helleseth {\em et al.} completely solved it in 2009 \cite{HeHoKhWaXi09}. For the second conjecture, the first part was proved in 2009, and the second part was proved in 2012 \cite{HeHoKhWaXi09,GHHK12}. Moreover, the dual function was also found in \cite{GHHK12} with the help of certain deep tools in algebraic number theory. In 2017, following the idea in \cite{GHHK12} with new observation, Hu {\em et al.} investigated the dual function of the Coulter-Matthews bent functions, and dug out a universal formula \cite{HZS17}. For two special cases, they found the dual function explicitly which yielded two classes of ternary bent functions not reported in the literature previously.

In this paper, using new combinatorial technique, for all cases of $k$ and $n$, we find a surprisingly explicit expression for the dual function of the Coulter-Matthews bent functions.
In particular, in some special cases, this expression can be improved much better. As a consequence, we find many classes of ternary bent functions which have not been reported in the literature previously. Such bent functions are not quadratic if $k>1$, and have $\left(\left(\frac{1+\sqrt{5}}{2}\right)^{w+1}-\right.$ $\left.\left(\frac{1-\sqrt{5}}{2}\right)^{w+1}\right)/\sqrt{5}$ or $\left(\left(\frac{1+\sqrt{5}}{2}\right)^{n-w+1}-\right.$ $\left.\left(\frac{1-\sqrt{5}}{2}\right)^{n-w+1}\right)/\sqrt{5}$ trace terms, where $0<w<n$ and $wk\equiv 1\ (\bmod\;n)$.
Among all cases, five special ones are especially interesting: for the case of $k=(n+1)/2$, the number of trace terms is $\left(\left(\frac{1+\sqrt{5}}{2}\right)^{n-1}-\right.$ $\left.\left(\frac{1-\sqrt{5}}{2}\right)^{n-1}\right)/\sqrt{5}$;
for the case of $k=n-1$, the number of trace terms is $\left(\left(\frac{1+\sqrt{5}}{2}\right)^n-\right.$ $\left.\left(\frac{1-\sqrt{5}}{2}\right)^n\right)/\sqrt{5}$;
for the case of $k=(n-1)/2$, the number of trace terms is $\left(\left(\frac{1+\sqrt{5}}{2}\right)^{n-1}-\right.$ $\left.\left(\frac{1-\sqrt{5}}{2}\right)^{n-1}\right)/\sqrt{5}$;
for the case of $(n, k)=(5t+4, 4t+3)$ or $(5t+1, 4t+1)$ with $t\geq 1$, the number of trace terms is 8;
and for the case of $(n, k)=(7t+6, 6t+5)$ or $(7t+1, 6t+1)$ with $t\geq 1$, the number of trace terms is 21.
As a consequence, we find new classes of ternary bent functions with only 8 or 21 trace terms. Several examples are also shown to verify our new findings.

This paper is organized as follows. Section \ref{sec_pre} lists some necessary notation and background. In Section \ref{sec_general}, we investigate
the general case, and give an explicit expression for all cases. Section \ref{sec_special} considers four special cases in detail, and shows better expressions. Finally, Section \ref{sec_con} concludes this paper.

\section{Preliminaries}\label{sec_pre}

\subsection{Cyclotomic Cosets Modulo $3^n-1$}

For any $0\leq s<3^n-1$, let $n_s$ be the smallest positive integer such that $s\equiv
s3^{n_s}(\mbox{mod }3^n-1)$. It is known that $n_s|n$. Then the cyclotomic coset $C_s$ modulo $3^n-1$ is defined to be the set
$$C_s=\{s, 3s, ..., 3^{n_s-1}s\}.$$
For simplicity, we may assume that $s$ is the
smallest integer in $C_s$. Then $s$ is called the coset leader of $C_s$ in this case. For example, for $n=2$, the cyclotomic cosets modulo 8 are:
$$C_0=\{0\}, C_1=\{1, 3\}, C_2=\{2, 6\}, C_4=\{4\}, C_5=\{5, 7\}.$$
Hence, $\{0, 1, 2, 4, 5\}$ are coset leaders modulo 8.

\begin{proposition}[Trace Representation \cite{GG05}]\label{prop_trace}
Any nonzero function $f(x)$ from $\mathbb{F}_{3^n}$ to
$\mathbb{F}_3$ can be represented as
$$f(x)=\sum_{k\in \Gamma(n)}Tr_1^{n_k}(F_kx^k)+F_{3^n-1}x^{3^n-1}, F_k\in \mathbb{F}_{3^{n_k}}, F_{3^n-1}\in \mathbb{F}_3$$
where $\Gamma(n)$ is the set consisting of all coset leaders modulo
$3^n-1$, $n_k|n$ is the size of the coset $C_k$, and $Tr_1^{n_k}(x)$
is the trace function from $\mathbb{F}_{3^{n_k}}$ to $\mathbb{F}_3$.
\end{proposition}

\subsection{The Ternary Modular Add-With-Carry Algorithm}

For any $0\leq k<3^n-1$, let $k=\sum_{i=0}^{n-1}k_i3^{i}$ be the
ternary representation of $k$, where $k_i\in\{0, 1, 2\}$ for any $0\leq i<n$.
Let $\mathrm{wt}(k)=\sum_{i=0}^{n-1}k_i$, and $\sigma(k)=\prod_{i=0}^{n-1}k_i!$. Moreover, for $j<0$ or $j\geq 3^n$, $\mathrm{wt}(j)$ and $\sigma(j)$ are used  to denote $\mathrm{wt}(\overline{j})$ and $\sigma(\overline{j})$ respectively, where $0\leq \overline{j}<3^n-1$
and $j\equiv \overline{j}\ (\bmod\;3^n-1)$.

For three integers $0\leq r, s, t<3^n-1$ satisfying $t\equiv r+s\mbox{ mod }3^n-1$, let $r=\sum_{i=0}^{n-1}r_i3^{i}$, $s=\sum_{i=0}^{n-1}s_i3^{i}$,
and $t=\sum_{i=0}^{n-1}t_i3^{i}$, where $0\leq r_i, s_i, t_i<3$ for any $0\leq i<n$. It is known that there is a unique integer sequence
$\overrightarrow{c}=c_0, c_1, ..., c_{n-1}$ with $c_i\in\{0, 1\}$ for any $0\leq i<n$ such that
$$t_i+3c_i=r_i+s_i+c_{i-1},\ 0\leq i\leq n-1,$$
where all subscripts take the value modulo $n$. From now on, for any $i$, the value $i$ modulo $n$ means the integer $\overline{i}$ satisfying
$0\leq \overline{i}<n$ and $\overline{i}\equiv i\ (\bmod\;n)$. Let $\mathrm{wt}(\overrightarrow{c})=c_0+c_1+...+c_{n-1}$. Then it follows that
$$\mathrm{wt}(r)+\mathrm{wt}(s)=\mathrm{wt}(t)+2\mathrm{wt}(\overrightarrow{c})\geq\mathrm{wt}(r+s).$$

\begin{proposition}[\cite{HZS17}]\label{prop_xy=}
With notation as above, $\mathrm{wt}(r)+\mathrm{wt}(s)=\mathrm{wt}(r+s)$ if and only if $r_i+s_i\leq 2$ for $0\leq i\leq n-1$,
and there exists $j$ with $0\leq j\leq n-1$ such that $r_j+s_j<2$.
In particular, $2\mathrm{wt}(r)=\mathrm{wt}(2r)$ if and only if $r_i\neq 2$ for $0\leq i\leq n-1$,
and $r\neq \frac{3^n-1}{2}$.
\end{proposition}

\begin{proposition}[\cite{HZS17}]\label{prop_xy+2}
With notation as above, $\mathrm{wt}(r)+\mathrm{wt}(s)=\mathrm{wt}(r+s)+2$ if and only if there exists $j$
with $0\leq j\leq n-1$ such that $r_j+s_j\geq 3$ and $r_{j+1}+s_{j+1}\leq 1$, and for $0\leq i\leq n-1$ with $i\neq j, j+1$, $r_i+s_i\leq 2$.
In particular, $2\mathrm{wt}(r)=\mathrm{wt}(2r)+2$ if and only if there exists $j$
with $0\leq j\leq n-1$ such that $r_j=2$ and $r_{j+1}=0$, and for $0\leq i\leq n-1$ with $i\neq j, j+1$, $r_i\leq 1$.
\end{proposition}

\subsection{The Dual of the Coulter-Matthews Bent Functions}

Using Stickelberger's theorem and the Teichm\"{u}ller character, a universal formula for the dual of the Coulter-Matthews bent functions has
been dug out in \cite{HZS17}. Note that Stickelberger's theorem has also played a significant role in the proof of some important conjectures and results \cite{CaChDo00,HoXi01,HSGH14,Mc72}.

Henceforth, let $d=(3^k+1)/2$ with $k$ odd and $\gcd(n, k)=1$, and $\eta$ be the quadratic character of $\mathbb{F}_{3^n}$ \cite{LN83}.

\begin{theorem}[\cite{HZS17}]\label{thm_main}
For any $a\in \mathbb{F}_{3^n}^*$ and $\lambda\in \mathbb{F}_{3^n}$, $\sum_{x\in \mathbb{F}_{3^n}}\omega_3^{Tr(ax^d+\lambda x)}=(-1)^{n+1}\eta(a)i^n\omega_3^{g(\lambda)}3^{n/2}$.
Moreover, we have $$g(\lambda)=\eta(a)\sum_{j: \mathrm{wt}(j)+\mathrm{wt}(-jd)=n+1}\sigma(j)\sigma(-jd)\left(\frac{a}{\lambda^d}\right)^j,$$
where $a/\lambda^d=0$ if $\lambda=0$.
\end{theorem}

For the case of $n=3t+2$ and $k=2t+1$ with $t\geq 1$, we have the following explicit expression for the dual function.

\begin{theorem}[\cite{HZS17}]\label{thm_3_term_1}
For any $a\in \mathbb{F}_{3^n}^*$ and $\lambda\in \mathbb{F}_{3^n}$, $\sum_{x\in \mathbb{F}_{3^n}}\omega_3^{Tr(ax^d+\lambda x)}=(-1)^{n+1}\eta(a)i^n\omega_3^{g(\lambda)}3^{n/2}$.
Moreover, we have
$$g(\lambda)=Tr\left(-\frac{\lambda^{3^{2t+2}+1}}{a^{3^{2t+2}-3^{t+1}+3}}-\frac{\lambda^{2\cdot3^{2t+1}+3^{t+1}+1}}{a^{3^{2t+2}+3^{t+1}+1}}+\frac{\lambda^2}{a^{-3^{2t+2}+3^{t+1}+3}}\right).$$
\end{theorem}

For the case of $n=3t+1$ and $k=2t+1$ with $t\geq 1$, we have another explicit expression for the dual function.

\begin{theorem}[\cite{HZS17}]\label{thm_3_term_2}
For any $a\in \mathbb{F}_{3^n}^*$ and $\lambda\in \mathbb{F}_{3^n}$, $\sum_{x\in \mathbb{F}_{3^n}}\omega_3^{Tr(ax^d+\lambda x)}=(-1)^{n+1}\eta(a)i^n\omega_3^{g(\lambda)}3^{n/2}$.
Moreover, we have
$$g(\lambda)=Tr\left(-\frac{\lambda^{3^{2t+1}+3^{t+1}+2}}{a^{3^{2t+1}+3^{t+1}+1}}-\frac{\lambda^{3^{2t}+1}}{a^{-3^{2t}+3^{t}+1}}+\frac{\lambda^{2}}{a^{-3^{2t+1}+3^{t+1}+1}}\right).$$
\end{theorem}

\section{The General Case}\label{sec_general}

For any $0\leq j<3^n-1$ with ternary representation $\sum_{i=0}^{n-1}j_i3^{i}$, i.e., $j_i\in\{0, 1, 2\}$ for $0\leq i<n$, we write $j$ in the form of $j_{n-1}j_{n-2}\cdots j_1j_0$ for simplicity. Moreover, for any $j$, let $\overline{j}$ be the residue of $j$ modulo $3^n-1$, i.e., $0\leq \overline{j}<3^n-1$ and $j\equiv \overline{j}\ (\bmod\;3^n-1)$. If $\overline{j}\neq 0$, then $\overline{-j}=\sum_{i=0}^{n-1}(2-j_i)3^{i}$, where $\overline{j}=j_{n-1}j_{n-2}\cdots j_1j_0$.

\begin{proposition}[\cite{HZS17}]\label{prop_n}
For any $0<j<3^n-1$, $\mathrm{wt}(j)+\mathrm{wt}(-jd)=n+1$ if and only if one of the following two conditions holds:

1) $\mathrm{wt}(j)+\mathrm{wt}(3^kj)=\mathrm{wt}((3^k+1)j)$ and $2\mathrm{wt}(-jd)=\mathrm{wt}(-(3^k+1)j)+2$;

2) $\mathrm{wt}(j)+\mathrm{wt}(3^kj)=\mathrm{wt}((3^k+1)j)+2$ and $2\mathrm{wt}(-jd)=\mathrm{wt}(-(3^k+1)j)$.
\end{proposition}

Let $$S_0=\{\ j\ |\ 0<j<3^n-1, \mathrm{wt}(j)+\mathrm{wt}(3^kj)=\mathrm{wt}((3^k+1)j), 2\mathrm{wt}(-jd)=\mathrm{wt}(-(3^k+1)j)+2\ \}$$ and
$$S_1=\{\ j\ |\ 0<j<3^n-1, \mathrm{wt}(j)+\mathrm{wt}(3^kj)=\mathrm{wt}((3^k+1)j)+2, 2\mathrm{wt}(-jd)=\mathrm{wt}(-(3^k+1)j)\ \}.$$
Then, by Proposition \ref{prop_n}, we have $S_0\bigcap S_1=\phi$ and
$$\{\ j\ |\ 0<j<3^n-1, \mathrm{wt}(j)+\mathrm{wt}(-jd)=n+1\ \}=S_0\bigcup S_1.$$

\subsection{The Set $S_0$}

Firstly, we have two lemmas from \cite{HZS17}.

\begin{lemma}[\cite{HZS17}]\label{lem_1_1}
For any $0<j<3^n-1$, let $h=h_{n-1}h_{n-2}\cdots h_1h_0=\overline{(3^k+1)j}$,
and $u=\overline{jd}$. If $\mathrm{wt}(j)+\mathrm{wt}(-jd)=n+1$ and $\mathrm{wt}(j)+\mathrm{wt}(3^kj)=\mathrm{wt}((3^k+1)j)$,
then there exists $0\leq a<n$ such that $h_a=h_{a+1}=1$, and for any $i\neq a, a+1$, $h_i=0$ or 2.
Besides, if $a=0$, then $u=(\frac{h}{2}+\frac{3^n-1}{2})(\bmod\;3^n-1)$.
\end{lemma}

\begin{lemma}[\cite{HZS17}]\label{lem_h/2}
For any $0<j<3^n-1$, let $h=\overline{(3^k+1)j}$, and $u=\overline{jd}$. If $\overline{3^kj}+j<3^n-1$, then $u=(\frac{h}{2}+\frac{3^n-1}{2})(\bmod\;3^n-1)$ if and only if $\sum_{i=n-k}^{n-1}j_i(\bmod\;2)=1$, where $j_{n-1}j_{n-2}\cdots j_1j_0$ is the ternary representation of $j$.
\end{lemma}

From now on, let $w$ be the integer satisfying $0<w<n$ and $wk\equiv 1\ (\bmod\;n)$, $\mathcal{A}=\{0, k, 2k, \cdots, (w-1)k\}(\bmod\;n)$, and $\mathcal{B}=\{wk, (w+1)k, \cdots, (n-1)k\}(\bmod\;n)$. Then $\mathcal{A}\cap \mathcal{B}=\phi$, and $\mathcal{A}\cup \mathcal{B}=\{0, 1, 2, \cdots, n-1\}$.

\begin{lemma}\label{lem_two_cases}
For any $0<j<3^n-1$, let $h=h_{n-1}h_{n-2}\cdots h_1h_0=\overline{(3^k+1)j}$. If $j\in S_0$ and $h_0=h_1=1$,
then there are two possible cases for $(j_1, j_0)$: $(1, 0)$ or $(0, 1)$.
\end{lemma}
{\bf Proof. }Let $j_{n-1}j_{n-2}\cdots j_1j_0$ be the ternary representation of $j$. Because $j\in S_0$, by Proposition \ref{prop_xy=}, we have
$j_i+j_{i-k}\leq 2$ for $0\leq i\leq n-1$. Moreover, since $h_0=h_1=1$, we get that $j_0+j_{-k}=j_1+j_{1-k}=1$. Thus, $\{j_0, j_{-k}\}=\{0, 1\}$, and $\{j_1, j_{1-k}\}=\{0, 1\}$. It follows that there are four cases for $(j_1, j_0)$: $(0, 0)$, $(1, 1)$, $(0, 1)$, or $(1, 0)$.

Suppose that $(j_1, j_0)=(0, 0)$. Then $(j_{1-k}, j_{-k})=(1, 1)$. By Lemma \ref{lem_1_1}, $h_{-k}=0$ or 2. On the other hand, $h_{-k}=j_{-k}+j_{-2k}=1+j_{-2k}$. Hence, $h_{-k}=2$, and $j_{-2k}=1$. Similarly, we have $j_{-ik}=1$ for $3\leq i\leq n-w$. However, $j_{-(n-w)k}=j_1$. It is a contradiction. Therefore, $(j_0, j_1)\not=(0, 0)$.

For the case of $(j_1, j_0)=(1, 1)$. The proof is similar. So we omit the details.

Consequently, there are only two possible cases for $(j_1, j_0)$: $(1, 0)$ or $(0, 1)$.\done

In the following, we have two lemmas regarding the ternary representation of $j\in S_0$.

\begin{lemma}\label{lem_10}
For any $0<j<3^n-1$, let $j=j_{n-1}j_{n-2}\cdots j_1j_0$, and $h=h_{n-1}h_{n-2}\cdots h_1h_0=\overline{(3^k+1)j}$. Suppose that $j\in S_0$ and $h_0=h_1=1$. If $(j_1, j_0)=(1, 0)$, then $j_{ik}=1$ for $w\leq i\leq n-1$, $j_{ik}=0$ for
$i\in \{0, w-1\}$, and $j_{ik}=0$ or 2 for $1\leq i\leq w-2$. Moreover, there is no consecutive 2's in the sequence $\{j_{ik}\}_{i=1}^{w-2}$ of length $w-2$.
\end{lemma}
{\bf Proof. }
Because $(j_1, j_0)=(1, 0)$, we have $(j_{1-k}, j_{-k})=(0, 1)$. Similar to the proof of Lemma \ref{lem_two_cases}, we get that $j_{-ik}=1$ for $1\leq i\leq n-w$, which means that $j_{ik}=1$ for $w\leq i\leq n-1$.

By Lemma \ref{lem_1_1}, $h_{1-k}=0$ or 2. On the other hand, $h_{1-k}=j_{1-k}+j_{1-2k}=j_{1-2k}$. Hence, $j_{1-2k}=0$ or 2. Now we consider $h_{1-2k}$. By Lemma \ref{lem_1_1}, $h_{1-2k}=0$ or 2. On the other hand, $h_{1-2k}=j_{1-2k}+j_{1-3k}$. If $j_{1-2k}=0$, then $j_{1-3k}=0$ or 2. If $j_{1-2k}=2$, then $j_{1-3k}=0$ by Proposition \ref{prop_xy=}. The same result holds for $j_{1-ik}=0$ with $4\leq i\leq w-1$. Hence, $j_{ik}=0$ or 2 for $1\leq i\leq w-2$, and there is no consecutive 2's in the sequence $\{j_{ik}\}_{i=1}^{w-2}$. \done

\begin{lemma}\label{lem_01}
For any $0<j<3^n-1$, let $j=j_{n-1}j_{n-2}\cdots j_1j_0$, and $h=h_{n-1}h_{n-2}\cdots h_1h_0=\overline{(3^k+1)j}$. Suppose that $j\in S_0$ and $h_0=h_1=1$. If $(j_1, j_0)=(0, 1)$, then $j_{ik}=1$ for $0\leq i\leq w-1$, $j_{ik}=0$ for
$i\in \{w, n-1\}$, and $j_{ik}=0$ or 2 for $w+1\leq i\leq n-2$. Moreover, there is no consecutive 2's in the sequence $\{j_{ik}\}_{i=w+1}^{n-2}$ of length $n-w-2$.
\end{lemma}
{\bf Proof. }The proof is similar to that of Lemma \ref{lem_10}. So we omit the details.
\done

In order to study the ternary representation of $j\in S_0$ further, we need another lemma.

\begin{lemma}\label{lem_AB}
$|\mathcal{A}\cap \{n-k, n-k+1, \cdots, n-1\}|+|\mathcal{B}\cap \{n-k, n-k+1, \cdots, n-1\}|=k$.
\end{lemma}
{\bf Proof. }
Because $\mathcal{A}\cap \mathcal{B}=\phi$ and $\mathcal{A}\cup \mathcal{B}=\{0, 1, 2, \cdots, n-1\}$, we have
\begin{eqnarray*}
\lefteqn{|\mathcal{A}\cap \{n-k, n-k+1, \cdots, n-1\}|+|\mathcal{B}\cap \{n-k, n-k+1, \cdots, n-1\}|}\\
&=&|(\mathcal{A}\cup \mathcal{B})\cap \{n-k, n-k+1, \cdots, n-1\}|\\
&=&|\{0, 1, 2, \cdots, n-1\}\cap \{n-k, n-k+1, \cdots, n-1\}|\\
&=&k.
\end{eqnarray*}
Therefore,
$|\mathcal{A}\cap \{n-k, n-k+1, \cdots, n-1\}|+|\mathcal{B}\cap \{n-k, n-k+1, \cdots, n-1\}|=k$.\done

\begin{lemma}\label{lem_one_case}
For any $0<j<3^n-1$, let $h=h_{n-1}h_{n-2}\cdots h_1h_0=\overline{(3^k+1)j}$. Suppose that $j\in S_0$ and $h_0=h_1=1$. If $|\mathcal{A}\cap \{n-k, n-k+1, \cdots, n-1\}|$ is even, then
$(j_1, j_0)=(1, 0)$; otherwise, $(j_1, j_0)=(0, 1)$.
\end{lemma}
{\bf Proof. }
By Lemma \ref{lem_two_cases}, there are two possible cases of $(j_1, j_0)$: $(1, 0)$ or $(0, 1)$. We need to show that only one case is correct.

1) Let $|\mathcal{A}\cap \{n-k, n-k+1, \cdots, n-1\}|$ be even. If $(j_1, j_0)=(0, 1)$, then $j_{ik}=1$ for $0\leq i\leq w-1$, and $j_{ik}\in \{0, 2\}$ for $w\leq i\leq n-1$ by Lemma \ref{lem_01}. It follows that
\begin{eqnarray*}
\sum_{i=n-k}^{n-1}j_i(\bmod\;2)&=&|\mathcal{A}\cap \{n-k, n-k+1, \cdots, n-1\}|(\bmod\;2)=0.
\end{eqnarray*}
By Lemmas \ref{lem_1_1} and \ref{lem_h/2}, it is a contradiction. Hence, $(j_1, j_0)=(1, 0)$.

2) Let $|\mathcal{A}\cap \{n-k, n-k+1, \cdots, n-1\}|$ be odd. If $(j_1, j_0)=(1, 0)$, then $j_{ik}=1$ for $w\leq i\leq n-1$, and $j_{ik}\in \{0, 2\}$ for $0\leq i\leq w-1$ by Lemma \ref{lem_01}. By Lemma \ref{lem_AB}, it follows that
\begin{eqnarray*}
\sum_{i=n-k}^{n-1}j_i(\bmod\;2)&=&|\mathcal{B}\cap \{n-k, n-k+1, \cdots, n-1\}|(\bmod\;2)\\
&=&(k-|\mathcal{A}\cap \{n-k, n-k+1, \cdots, n-1\}|)(\bmod\;2)\\
&=&0.
\end{eqnarray*}
By Lemmas \ref{lem_1_1} and \ref{lem_h/2}, it is also a contradiction. Hence, $(j_1, j_0)=(0, 1)$.
\done

In order to study the cardinality of $S_0$, the number of a class of special sequences is necessary for the investigation.
Let $\{x_i\}_{i=1}^f$ be a sequence of length $f$, where $x_i=0$ or 2 for $1\leq i\leq f$. Moreover, there is no consecutive 2's in the sequence.
Let $N(f)$ denote the number of $\{x_i\}_{i=1}^f$. In particular, let $N(0)=1$.

\begin{lemma}\label{lem_number_sequence}
With notation as above, $N(f)=\left(\left(\frac{1+\sqrt{5}}{2}\right)^{f+2}-\left(\frac{1-\sqrt{5}}{2}\right)^{f+2}\right)/\sqrt{5}$ for any $f\geq0$.
\end{lemma}
{\bf Proof. }
One may check that $N(1)=2$, and $N(f)=N(f-1)+N(f-2)$ for any $f\geq 2$. Therefore, $N(f)=\left(\left(\frac{1+\sqrt{5}}{2}\right)^{f+2}-\left(\frac{1-\sqrt{5}}{2}\right)^{f+2}\right)/\sqrt{5}$.
\done

\begin{lemma}\label{lem_coset}
For any $0<j<3^n-1$, let $h=h_{n-1}h_{n-2}\cdots h_1h_0=\overline{(3^k+1)j}$. Suppose that $j\in S_0$ and $h_0=h_1=1$. Let $C_j$ be the coset containing $j$. Then $C_j\subset S_0$, and $|C_j|=n$.
\end{lemma}
{\bf Proof. }
One may check that $C_j\subset S_0$. Suppose that $|C_j|=m<n$. Let $\widetilde{j}=3^mj(\mbox{mod }3^n-1)$, and $\widetilde{h}=\widetilde{h}_{n-1}\widetilde{h}_{n-2}\cdots \widetilde{h}_1\widetilde{h}_0=\overline{(3^k+1)\widetilde{j}}$. Then, $\widetilde{j}=j$, and
$\widetilde{h}=h$. However, $\widetilde{h}=3^mh(\mbox{mod }3^n-1)$. Hence, by Lemma \ref{lem_1_1}, $(\widetilde{h}_0, \widetilde{h}_1)\not=(1, 1)$. It is a contradiction. Therefore, $|C_j|=n$.
\done

Let
\begin{eqnarray}
U_0&=&\{\ j\ |\ j_{ik}=1\mbox{ for }w\leq i\leq n-1, j_{ik}=0\mbox{ for
}i\in \{0, w-1\}, \nonumber\\
&&\ \ \ j_{ik}\in\{0, 2\}\mbox{ for }1\leq i\leq w-2,\mbox{ and no consecutive 2's in }\{j_{ik}\}_{i=1}^{w-2}\ \},\label{eqn_U0}
\end{eqnarray}
and
\begin{eqnarray}
U_1&=&\{\ j\ |\ j_{ik}=1\mbox{ for }0\leq i\leq w-1, j_{ik}=0\mbox{ for
}i\in \{w, n-1\},\nonumber\\
&&\ \ \ j_{ik}\in\{0, 2\}\mbox{ for }w+1\leq i\leq n-2,\mbox{ and no consecutive 2's in }\{j_{ik}\}_{i=w+1}^{n-2}\ \}.\label{eqn_U1}
\end{eqnarray}
By Lemma \ref{lem_number_sequence},
we know that $$|U_0|=\left(\left(\frac{1+\sqrt{5}}{2}\right)^{w}-\left(\frac{1-\sqrt{5}}{2}\right)^{w}\right)/\sqrt{5}\mbox{ and }
|U_1|=\left(\left(\frac{1+\sqrt{5}}{2}\right)^{n-w}-\left(\frac{1-\sqrt{5}}{2}\right)^{n-w}\right)/\sqrt{5}.$$

By Lemmas \ref{lem_10}, \ref{lem_01}, \ref{lem_one_case}, \ref{lem_number_sequence}, and \ref{lem_coset}, we obtain the following theorem.

\begin{theorem}\label{thm_S0}
With notation as above, if $|\mathcal{A}\cap \{n-k, n-k+1, \cdots, n-1\}|$ is even, then $S_0=\bigcup_{j\in U_0}C_j$,
and $|S_0|=n\left(\left(\frac{1+\sqrt{5}}{2}\right)^{w}-\left(\frac{1-\sqrt{5}}{2}\right)^{w}\right)/\sqrt{5}$;
otherwise, $S_0=\bigcup_{j\in U_1}C_j$, and $|S_0|=n\left(\left(\frac{1+\sqrt{5}}{2}\right)^{n-w}-\right.$ $\left.\left(\frac{1-\sqrt{5}}{2}\right)^{n-w}\right)/\sqrt{5}$.
\end{theorem}

\begin{example}\label{example_1}
Let $n=8$, and $k=7$. Then $w=7$, and $|\mathcal{A}\cap \{n-k, n-k+1, \cdots, n-1\}|=6$. By Theorem \ref{thm_S0}, $S_0=\bigcup_{j\in U_0}C_j$, and
$|S_0|=104$. By computer program, one may check that $U_0=\{00000010, 00002010, 00020010, 00200010, 00202010, 02000010, 02002010, 02020010, 20000010,
20002010,$\\$20020010, 20200010, 20202010\}$.
\end{example}

\begin{example}\label{example_2}
Let $n=9$, and $k=7$. Then $w=4$, and $|\mathcal{A}\cap \{n-k, n-k+1, \cdots, n-1\}|=3$. By Theorem \ref{thm_S0}, $S_0=\bigcup_{j\in U_1}C_j$, and
$|S_0|=45$. By computer program, one may check that $U_1=\{010101001, 010121001, 012101001, 210101001, 210121001\}$.
\end{example}

\subsection{The Set $S_1$}

Firstly, we have one lemma from \cite{HZS17}.

\begin{lemma}[\cite{HZS17}]\label{lem_1_2}
For any $0<j<3^n-1$, let $h=h_{n-1}h_{n-2}\cdots h_1h_0=\overline{(3^k+1)j}$, and $u=\overline{jd}$. If $\mathrm{wt}(j)+\mathrm{wt}(-jd)=n+1$ and $\mathrm{wt}(j)+\mathrm{wt}(3^kj)=\mathrm{wt}((3^k+1)j)+2$, then
$h_i=0$ or 2 for any $0\leq i<n$, and there exists $0\leq a<n$ such that $\{j_a, j_{-k+a}\}=\{1, 2\}$ and $\{j_{a+1}, j_{-k+a+1}\}=\{0, 1\}$, where $j_{n-1}j_{n-2}\cdots j_1j_0$ is the ternary representation of $j$. Besides, if $a=0$, then $u=(\frac{h}{2}+\frac{3^n-1}{2})(\bmod\;3^n-1)$.
\end{lemma}

\begin{lemma}\label{lem_two_cases_2}
For any $0<j<3^n-1$, let $j=j_{n-1}j_{n-2}\cdots j_1j_0$. Suppose that $j\in S_1$. If $\{j_0, j_{-k}\}=\{1, 2\}$ and $\{j_{1}, j_{-k+1}\}=\{0, 1\}$, then
there are two possible cases for $(j_1, j_0)$: $(1, 2)$ or $(0, 1)$.
\end{lemma}
{\bf Proof. }
Let $h=h_{n-1}h_{n-2}\cdots h_1h_0=\overline{(3^k+1)j}$. Because $j\in S_1$, by Lemma \ref{lem_1_2}, we have $h_i=0$ or 2 for any $0\leq i<n$.
Moreover, because $j_0+j_{-k}=3$, by Proposition \ref{prop_xy+2}, $j_i+j_{i-k}\leq 2$ for $0\leq i<n$ with $i\neq 0, 1$.

Because $\{j_0, j_{-k}\}=\{1, 2\}$ and $\{j_{1}, j_{-k+1}\}=\{0, 1\}$, there are four possible cases for $(j_1, j_0)$: $(1, 1), (1, 2), (0, 1)$, or $(0, 2)$.

Suppose that $(j_1, j_0)=(1, 1)$. Then $(j_{1-k}, j_{-k})=(0, 2)$. Let us consider $h_k=j_k+j_0=j_k+1$. Because $h_i\in\{0, 2\}$ and $j_k+j_0\leq 2$, we get
$j_k=1$. Similarly, $j_{ik}=1$ for $2\leq i\leq w-1$. However, $j_{(w-1)k}=j_{1-k}=0$. It is a contradiction. Hence, $(j_1, j_0)\not=(1, 1)$.

Suppose that $(j_1, j_0)=(0, 2)$. Then $(j_{1-k}, j_{-k})=(1, 1)$. Similarly, we have $j_{ik}=1$ for $0\leq i\leq w-1$.  However, $j_0=2$. It is a contradiction. Hence, $(j_1, j_0)\not=(0, 2)$.

Consequently, there are only two possible cases for $(j_1, j_0)$: $(1, 2)$ or $(0, 1)$.\done

\begin{lemma}\label{lem_12}
For any $0<j<3^n-1$, let $j=j_{n-1}j_{n-2}\cdots j_1j_0$. Suppose that $j\in S_1$, $\{j_0, j_{-k}\}=\{1, 2\}$, and $\{j_{1}, j_{-k+1}\}=\{0, 1\}$.
If $(j_1, j_0)=(1, 2)$, then $j_{ik}=1$ for $w\leq i\leq n-1$, $j_{ik}=2$ for $i=0$, $j_{ik}=0$ for
$i\in \{1, w-1\}$, and $j_{ik}=0$ or 2 for $2\leq i\leq w-2$. Moreover, there is no consecutive 2's in the sequence $\{j_{ik}\}_{i=2}^{w-2}$ of length $w-3$.
\end{lemma}
{\bf Proof. }
Let $h=h_{n-1}h_{n-2}\cdots h_1h_0=\overline{(3^k+1)j}$. Similar to the proof of Lemma \ref{lem_two_cases_2}, we have $h_i=0$ or 2 for any $0\leq i<n$,
and $j_i+j_{i-k}\leq 2$ for $1<i<n$.

If $(j_1, j_0)=(1, 2)$, then $(j_{1-k}, j_{-k})=(0, 1)$. Similar to the proof of Lemma \ref{lem_two_cases_2},
we get that $j_{ik}=1$ for $w\leq i\leq n-1$.

Note that $h_k=j_k+j_0=j_k+2$. On the other hand, $h_k=0$ or 2, and $j_k+j_0\leq 2$. Hence, $j_k=0$. Let us consider $h_{2k}$. Since
$h_{2k}=j_{2k}+j_k=j_{2k}$, we know that $j_{2k}=0$ or 2. Similarly, if $j_{2k}=0$, then $j_{3k}=0$ or 2; otherwise, $j_{3k}=0$. The same result
holds for $j_{ik}$ with $4\leq i\leq w-2$. Hence, $j_{ik}=0$ or 2 for $2\leq i\leq w-2$, and there is no consecutive 2's in the sequence $\{j_{ik}\}_{i=2}^{w-2}$.
\done

\begin{lemma}\label{lem_01_2}
For any $0<j<3^n-1$, let $j=j_{n-1}j_{n-2}\cdots j_1j_0$. Suppose that $j\in S_1$, $\{j_0, j_{-k}\}=\{1, 2\}$, and $\{j_{1}, j_{-k+1}\}=\{0, 1\}$.
If $(j_1, j_0)=(0, 1)$, then $j_{ik}=1$ for $0\leq i\leq w-1$, $j_{ik}=0$ for
$i\in \{w, n-2\}$, $j_{ik}=0$ or 2 for $w+1\leq i\leq n-3$, and $j_{ik}=2$ for $i=n-1$. Moreover, there is no consecutive 2's in the sequence $\{j_{ik}\}_{i=w+1}^{n-3}$ of length $n-w-3$.
\end{lemma}
{\bf Proof. }The proof is similar to that of Lemma \ref{lem_12}. So we omit the details.
\done

\begin{lemma}\label{lem_one_case_2}
For any $0<j<3^n-1$, let $j=j_{n-1}j_{n-2}\cdots j_1j_0$. Suppose that $j\in S_1$, $\{j_0, j_{-k}\}=\{1, 2\}$, and $\{j_{1}, j_{-k+1}\}=\{0, 1\}$.
If $|\mathcal{A}\cap \{n-k, n-k+1, \cdots, n-1\}|$ is even, then $(j_1, j_0)=(1, 2)$; otherwise, $(j_1, j_0)=(0, 1)$.
\end{lemma}
{\bf Proof. }The proof is similar to that of Lemma \ref{lem_one_case}. So we omit the details.
\done

\begin{lemma}\label{lem_coset_2}
For any $0<j<3^n-1$, let $j=j_{n-1}j_{n-2}\cdots j_1j_0$. Suppose that $j\in S_1$, $\{j_0, j_{-k}\}=\{1, 2\}$, and $\{j_{1}, j_{-k+1}\}=\{0, 1\}$.
Let $C_j$ be the coset containing $j$. Then $C_j\subset S_1$, and $|C_j|=n$.
\end{lemma}
{\bf Proof. }One may check that $C_j\subset S_1$. Suppose that $|C_j|=m<n$. Because $j\in S_1$ and $j_0+j_{-k}=3$, by Proposition \ref{prop_xy+2}, $j_i+j_{i-k}\leq 2$ for $0<i<n$.
Let $\widetilde{j}=3^mj(\mbox{mod }3^n-1)$. Then $\widetilde{j}=j$, which means that $\{j_{-m}, j_{-k-m}\}=\{j_0, j_{-k}\}$. It follows that
$j_{-m}+j_{-k-m}=j_0+j_{-k}=3$. It is a contradiction. Therefore, $|C_j|=n$.
\done

Let
\begin{eqnarray}
V_0&=&\{\ j\ |\ j_{ik}=1\mbox{ for }w\leq i\leq n-1, j_{ik}=2\mbox{ for }i=0, j_{ik}=0\mbox{ for
}i\in \{1, w-1\}, \nonumber\\
&&\ \ \ j_{ik}\in\{0, 2\}\mbox{ for }2\leq i\leq w-2,\mbox{ and no consecutive 2's in }\{j_{ik}\}_{i=2}^{w-2}\ \},\label{eqn_V0}
\end{eqnarray}
and
\begin{eqnarray}
V_1&=&\{\ j\ |\ j_{ik}=1\mbox{ for }0\leq i\leq w-1, j_{ik}=0\mbox{ for
}i\in \{w, n-2\}, j_{ik}\in\{0, 2\}\mbox{ for }w+1\leq i\leq n-3,\nonumber\\
&&\ \ \ j_{ik}=2\mbox{ for }i=n-1,\mbox{ and no consecutive 2's in }\{j_{ik}\}_{i=w+1}^{n-3}\ \}.\label{eqn_V1}
\end{eqnarray}
By Lemma \ref{lem_number_sequence},
it follows that $|V_0|=\left(\left(\frac{1+\sqrt{5}}{2}\right)^{w-1}-\left(\frac{1-\sqrt{5}}{2}\right)^{w-1}\right)/\sqrt{5}$,
and $|V_1|=\left(\left(\frac{1+\sqrt{5}}{2}\right)^{n-w-1}-\right.$ $\left.\left(\frac{1-\sqrt{5}}{2}\right)^{n-w-1}\right)/\sqrt{5}$.

By Lemmas \ref{lem_number_sequence}, \ref{lem_12}, \ref{lem_01_2}, \ref{lem_one_case_2}, and \ref{lem_coset_2}, we obtain the following theorem.

\begin{theorem}\label{thm_S1}
With notation as above, if $|\mathcal{A}\cap \{n-k, n-k+1, \cdots, n-1\}|$ is even, then $S_1=\bigcup_{j\in V_0}C_j$,
and $|S_1|=n\left(\left(\frac{1+\sqrt{5}}{2}\right)^{w-1}-\left(\frac{1-\sqrt{5}}{2}\right)^{w-1}\right)/\sqrt{5}$;
otherwise, $S_1=\bigcup_{j\in V_1}C_j$, and $|S_1|=n\left(\left(\frac{1+\sqrt{5}}{2}\right)^{n-w-1}-\right.$ $\left.\left(\frac{1-\sqrt{5}}{2}\right)^{n-w-1}\right)/\sqrt{5}$.
\end{theorem}

\begin{example}\label{example_3}
Let $n=8$, and $k=7$. Then $w=7$, and $|\mathcal{A}\cap \{n-k, n-k+1, \cdots, n-1\}|=6$. By Theorem \ref{thm_S1}, $S_1=\bigcup_{j\in V_0}C_j$, and
$|S_1|=64$. By computer program, one may check that $V_0=\{00000012, 00002012, 00020012, 00200012, 00202012, 02000012, 02002012, 02020012\}$.
\end{example}

\begin{example}\label{example_4}
Let $n=9$, and $k=7$. Then $w=4$, and $|\mathcal{A}\cap \{n-k, n-k+1, \cdots, n-1\}|=3$. By Theorem \ref{thm_S1}, $S_1=\bigcup_{j\in V_1}C_j$, and
$|S_1|=27$. By computer program, one may check that $V_1=\{010101201, 012101201, 210101201\}$.
\end{example}

\subsection{The Value of $\sigma(j)\sigma(-jd)$ for any $j\in S_0\bigcup S_1$}

For any $0<j<3^n-1$, let $N_2(j)$ denote the number of $2$'s in the ternary representation of $j$. Then $\sigma(j)=2^{N_2(j)}$.
In order to study $\sigma(j)\sigma(-jd)$, we have two lemmas listed below.

\begin{lemma}\label{lem_S0}
If $j\in S_0$, then $\sigma(j)\sigma(-jd)=2^{N_2(j)+1}$.
\end{lemma}
{\bf Proof. }Because $S_0=\sum_{j\in U_0}C_j$ or $\sum_{j\in U_1}C_j$, where $C_j$ is the coset containing $j$, we only need to consider the case of $j\in U_0$ or $j\in U_1$.

Suppose that $j\in U_0$.
Let $h=h_{n-1}h_{n-2}\cdots h_1h_0=\overline{(3^k+1)j}$, and $u=u_{n-1}u_{n-2}\cdots u_1u_0=\overline{jd}$.
Then $h_1=h_0=1$, and $h_i=0$ or 2 for $1<i<n$. By Lemma \ref{lem_1_1}, $u=(\frac{h}{2}+\frac{3^n-1}{2})(\bmod\;3^n-1)$.
Therefore, for $1<i<n$, $u_i=h_i/2+1=1$ or 2. Besides, $u_1=2$, and $u_0=0$. Let $v=v_{n-1}v_{n-2}\cdots v_1v_0=\overline{-jd}$.
Then $v=\overline{-u}$. It follows that $v_0=2$, $v_1=0$, and $v_i=0$ or 1 for $1<i<n$. Thus, we have $\sigma(-jd)=\sigma(v)=2$.

For the case of $j\in U_1$, the proof is the same. So we omit the details.
\done

\begin{lemma}\label{lem_S1}
If $j\in S_1$, then $\sigma(j)\sigma(-jd)=2^{N_2(j)}$.
\end{lemma}
{\bf Proof. }Because $S_1=\sum_{j\in V_0}C_j$ or $\sum_{j\in V_1}C_j$, where $C_j$ is the coset containing $j$, we only need to consider the case of $j\in V_0$ or $j\in V_1$.

Suppose that $j\in V_0$.
Let $h=h_{n-1}h_{n-2}\cdots h_1h_0=\overline{(3^k+1)j}$, and $u=u_{n-1}u_{n-2}\cdots u_1u_0=\overline{jd}$.
Then $h_i=0$ or 2 for $0\leq i<n$. By Lemma \ref{lem_1_1}, $u=(\frac{h}{2}+\frac{3^n-1}{2})(\bmod\;3^n-1)$.
Therefore, $u_i=h_i/2+1=1$ or 2 with $0\leq i<n$. Let $v=v_{n-1}v_{n-2}\cdots v_1v_0=\overline{-jd}$.
Then $v=\overline{-u}$. It follows that $v_i=0$ or 1 for $0\leq i<n$. Thus, we have $\sigma(-jd)=\sigma(v)=1$.

For the case of $j\in V_1$, the proof is the same. So we omit the details.
\done

The following theorem is the main result of this section.
\begin{theorem}\label{thm_main_2}
Let $g(x)$ be the dual function of the Coulter-Matthews bent function $Tr(ax^{d})$, where $a\in\mathbb{F}_{3^n}^{*}$. On the trace representation of $g(x)$, there are two cases as follows.

1) If $|\mathcal{A}\cap \{n-k, n-k+1, \cdots, n-1\}|$ is even, then
$$g(x)=\sum_{j\in U_0}Tr\left((-1)^{N_2(j)+1}\eta(a)a^jx^{-jd}\right)+\sum_{j\in V_0}Tr\left((-1)^{N_2(j)}\eta(a)a^jx^{-jd}\right),$$
where $x^{-jd}=0$ if $x=0$, $U_0$ is defined by (\ref{eqn_U0}), and $V_0$ is defined by (\ref{eqn_V0}). Besides, there are $\left(\left(\frac{1+\sqrt{5}}{2}\right)^{w+1}-\right.$ $\left.\left(\frac{1-\sqrt{5}}{2}\right)^{w+1}\right)/\sqrt{5}$ trace terms.

2) If $|\mathcal{A}\cap \{n-k, n-k+1, \cdots, n-1\}|$ is odd, then
$$g(x)=\sum_{j\in U_1}Tr\left((-1)^{N_2(j)+1}\eta(a)a^jx^{-jd}\right)+\sum_{j\in V_1}Tr\left((-1)^{N_2(j)}\eta(a)a^jx^{-jd}\right),$$
where $x^{-jd}=0$ if $x=0$, $U_1$ is defined by (\ref{eqn_U1}), and $V_1$ is defined by (\ref{eqn_V1}). Besides, there are $\left(\left(\frac{1+\sqrt{5}}{2}\right)^{n-w+1}-\right.$ $\left.\left(\frac{1-\sqrt{5}}{2}\right)^{n-w+1}\right)/\sqrt{5}$ trace terms.
\end{theorem}
{\bf Proof. }By Theorem \ref{thm_main} and Proposition \ref{prop_n}, we have
\begin{eqnarray*}
g(x)&=&\eta(a)\sum_{j: \mathrm{wt}(j)+\mathrm{wt}(-jd)=n+1}\sigma(j)\sigma(-jd)\left(\frac{a}{x^d}\right)^j\\
&=&\eta(a)\sum_{j\in S_0\bigcup S_1}\sigma(j)\sigma(-jd)\left(\frac{a}{x^d}\right)^j\\
&=&\eta(a)\sum_{j\in S_0\bigcup S_1}\sigma(j)\sigma(-jd)a^jx^{-jd}.
\end{eqnarray*}
By Lemmas \ref{lem_S0} and \ref{lem_S1}, it holds that
\begin{eqnarray*}
g(x)&=&\eta(a)\sum_{j\in S_0}2^{N_2(j)+1}a^jx^{-jd}+\eta(a)\sum_{j\in S_1}2^{N_2(j)}a^jx^{-jd}\\
&=&\sum_{j\in S_0}(-1)^{N_2(j)+1}\eta(a)a^jx^{-jd}+\sum_{j\in S_1}(-1)^{N_2(j)}\eta(a)a^jx^{-jd}.
\end{eqnarray*}

1) If $|\mathcal{A}\cap \{n-k, n-k+1, \cdots, n-1\}|$ is even, by Theorems \ref{thm_S0} and \ref{thm_S1}, it follows that
\begin{eqnarray*}
g(x)&=&\sum_{j\in U_0}Tr((-1)^{N_2(j)+1}\eta(a)a^jx^{-jd})+\sum_{j\in V_0}Tr((-1)^{N_2(j)}\eta(a)a^jx^{-jd}).
\end{eqnarray*}
There are $|U_0|+|V_0|=\left(\left(\frac{1+\sqrt{5}}{2}\right)^{w+1}-\left(\frac{1-\sqrt{5}}{2}\right)^{w+1}\right)/\sqrt{5}$ trace terms.

2) If $|\mathcal{A}\cap \{n-k, n-k+1, \cdots, n-1\}|$ is odd, by Theorems \ref{thm_S0} and \ref{thm_S1}, it follows that
\begin{eqnarray*}
g(x)&=&\sum_{j\in U_1}Tr((-1)^{N_2(j)+1}\eta(a)a^jx^{-jd})+\sum_{j\in V_1}Tr((-1)^{N_2(j)}\eta(a)a^jx^{-jd}).
\end{eqnarray*}
There are $|U_1|+|V_1|=\left(\left(\frac{1+\sqrt{5}}{2}\right)^{n-w+1}-\left(\frac{1-\sqrt{5}}{2}\right)^{n-w+1}\right)/\sqrt{5}$ trace terms.
\done

\begin{example}\label{example_5}
Let $n=8$, and $k=7$. Then $d=1094$, $w=7$, and $|\mathcal{A}\cap \{n-k, n-k+1, \cdots, n-1\}|=6$. By Examples \ref{example_1} and \ref{example_3},
$U_0=\{00000010, 00002010, 00020010, 00200010, 00202010, 02000010, 02002010, $ $02020010, 20000010,
20002010, 20020010, 20200010, 20202010\}$, and $V_0=\{00000012, 00002012, 00020012, $ $00200012, 00202012, 02000012, 02002012, 02020012\}$.
By Theorem \ref{thm_main_2}, for any $a\in \mathbb{F}_{3^8}^{*}$, the dual function $g(x)$ of the Coulter-Matthews bent function $Tr(ax^{1094})$
over $\mathbb{F}_{3^8}$ is given by
$$g(x)=\sum_{j\in U_0}Tr((-1)^{N_2(j)+1}\eta(a)a^jx^{-1094j})+\sum_{j\in V_0}Tr((-1)^{N_2(j)}\eta(a)a^jx^{-1094j}).$$
Totally, there are 21 trace terms.
\end{example}

\begin{example}\label{example_6}
Let $n=9$, and $k=7$. Then $d=1094$, $w=4$, and $|\mathcal{A}\cap \{n-k, n-k+1, \cdots, n-1\}|=3$. By Examples \ref{example_2} and \ref{example_4},
$U_1=\{010101001, 010121001, 012101001, 210101001, 210121001\}$, and $V_1=\{010101201, 012101201, 210101201\}$. By Theorem \ref{thm_main_2},
for any $a\in \mathbb{F}_{3^9}^{*}$, the dual function $g(x)$ of the Coulter-Matthews bent function $Tr(ax^{1094})$
over $\mathbb{F}_{3^9}$ is given by
$$g(x)=\sum_{j\in U_1}Tr((-1)^{N_2(j)+1}\eta(a)a^jx^{-1094j})+\sum_{j\in V_1}Tr((-1)^{N_2(j)}\eta(a)a^jx^{-1094j}).$$
Totally, there are 8 trace terms.
\end{example}

\begin{theorem}
With notation as in Theorem \ref{thm_main_2}, the algebraic degree of $g(x)$ is $w+1$ if $|\mathcal{A}\cap \{n-k, n-k+1, \cdots, n-1\}|$ is even, or $n+1-w$ if $|\mathcal{A}\cap \{n-k, n-k+1, \cdots, n-1\}|$ is odd. Moreover, if $k>1$, then the algebraic degree of $g(x)$ is bigger than 2.
\end{theorem}
{\bf Proof. }

1) If $|\mathcal{A}\cap \{n-k, n-k+1, \cdots, n-1\}|$ is even, by Theorem \ref{thm_main_2}, the algebraic degree of $g(x)$ is equal to
$$\max_{j\in U_0\bigcup V_0} \{\mathrm{wt}(-jd)\}=\max_{j\in U_0\bigcup V_0} \{n+1-\mathrm{wt}(j)\}=n+1-\min_{j\in U_0\bigcup V_0} \{\mathrm{wt}(j)\}.$$
By (\ref{eqn_U0}) and (\ref{eqn_V0}), $\min_{j\in U_0\bigcup V_0} \{\mathrm{wt}(j)\}=n-w$. Hence, $\max_{j\in U_0\bigcup V_0} \{\mathrm{wt}(-jd)\}=w+1\geq 2$.
If $w+1=2$, then $w=1$, which means that $k=1$. Therefore, in this case, the algebraic degree of $g(x)$ is bigger than 2 if $k>1$.

2) If $|\mathcal{A}\cap \{n-k, n-k+1, \cdots, n-1\}|$ is odd, by Theorem \ref{thm_main_2}, the algebraic degree of $g(x)$ is equal to
$$\max_{j\in U_1\bigcup V_1} \{\mathrm{wt}(-jd)\}=\max_{j\in U_1\bigcup V_1} \{n+1-\mathrm{wt}(j)\}=n+1-\min_{j\in U_1\bigcup V_1} \{\mathrm{wt}(j)\}.$$
By (\ref{eqn_U1}) and (\ref{eqn_V1}), $\min_{j\in U_1\bigcup V_1} \{\mathrm{wt}(j)\}=w$. Hence, $\max_{j\in U_1\bigcup V_1} \{\mathrm{wt}(-jd)\}=n+1-w\geq 2$.
If $n+1-w=2$, then $w=n-1$, which means that $k=n-1$. However, if $k=n-1$, then $|\mathcal{A}\cap \{n-k, n-k+1, \cdots, n-1\}|=k-1$. It follows that $|\mathcal{A}\cap \{n-k, n-k+1, \cdots, n-1\}|$ is even, and we get a contradiction. Therefore, in this case, the algebraic degree of $g(x)$ is also bigger than 2. \done

\section{Some Special Cases}\label{sec_special}

In this section, we investigate some special cases further, and obtain more kinds of bent functions with explicit trace representation.
In particular, we find new classes of bent functions with only 8 or 21 trace terms.

\subsection{The Case of $k|n+1$ and $k>1$}\label{subsec_n+1}

\begin{lemma}\label{lem_k_n+1}
If $k|n+1$ and $k>1$, then $w=(n+1)/k$. Moreover, we have $|\mathcal{A}\cap \{n-k, n-k+1, \cdots, n-1\}|=1$.
\end{lemma}
{\bf Proof. }Let $x=(n+1)/k$. Then $0<x<n$, and $xk\equiv 1\ (\bmod\;n)$. Hence, $w=x$. As a consequence,
$$\mathcal{A}=\{0, k, 2k, \cdots, (w-1)k\}(\bmod\;n)=\{0, k, 2k, \cdots, n+1-2k, n+1-k\}.$$
It follows that $\mathcal{A}\cap \{n-k, n-k+1, \cdots, n-1\}=\{n-k+1\}$, which means that $|\mathcal{A}\cap \{n-k, n-k+1, \cdots, n-1\}|=1$.
\done

\begin{theorem}\label{thm_k_n+1}
Let $g(x)$ be the dual function of the Coulter-Matthews bent function $Tr(ax^{d})$, where $a\in\mathbb{F}_{3^n}^{*}$.
If $k|n+1$ and $k>1$, then
$$g(x)=\sum_{j\in U_1}Tr\left((-1)^{N_2(j)+1}\eta(a)a^jx^{-jd}\right)+\sum_{j\in V_1}Tr\left((-1)^{N_2(j)}\eta(a)a^jx^{-jd}\right),$$
where $x^{-jd}=0$ if $x=0$, $U_1$ is defined by (\ref{eqn_U1}), and $V_1$ is defined by (\ref{eqn_V1}). Besides, there are $$\left(\left(\frac{1+\sqrt{5}}{2}\right)^{(n+1)(k-1)/k}-\left(\frac{1-\sqrt{5}}{2}\right)^{(n+1)(k-1)/k}\right)/\sqrt{5}$$ trace terms.
\end{theorem}
{\bf Proof. }
By Lemma \ref{lem_k_n+1}, $w=(n+1)/k$, and $|\mathcal{A}\cap \{n-k, n-k+1, \cdots, n-1\}|$ is odd.
By Theorem \ref{thm_main_2}, the result follows.
\done

\begin{remark}
With notation as in Theorem \ref{thm_k_n+1}, the following case is especially interesting: if $k=(n+1)/2$, then there are $\left(\left(\frac{1+\sqrt{5}}{2}\right)^{n-1}-\right.$ $\left.\left(\frac{1-\sqrt{5}}{2}\right)^{n-1}\right)/\sqrt{5}$ trace terms.
\end{remark}

\begin{example}
Let $n=9$, and $k=5$. Then $k=(n+1)/2$, and $d=122$. By (\ref{eqn_U1}), $U_1=\{000100001, 0001002$ $01, 000102001, 000102201, 002100001, 002102001, 020100001, 022100001, 200100001, 200100201, 2021000$\\ $01, 220100001, 222100001\}$, and by (\ref{eqn_V1}), $V_1=\{000120001, 000120201, 000122001, 000122201, 002120001, $ $002122001, 020120001, 022120001\}$. By Theorem \ref{thm_k_n+1}, for any $a\in \mathbb{F}_{3^9}^{*}$, the dual function $g(x)$ of the Coulter-Matthews bent function $Tr(ax^{122})$ over $\mathbb{F}_{3^9}$ is given by
$$g(x)=\sum_{j\in U_1}Tr((-1)^{N_2(j)+1}\eta(a)a^jx^{-122j})+\sum_{j\in V_1}Tr((-1)^{N_2(j)}\eta(a)a^jx^{-122j}).$$
Totally, there are 21 trace terms.
\end{example}

\subsection{The Case of $k|n-1$ and $k>1$}\label{subsec_n-1}

\begin{lemma}\label{lem_k_n-1}
If $k|n-1$ and $k>1$, then $w=n-(n-1)/k$. Moreover, we have $|\mathcal{A}\cap \{n-k, n-k+1, \cdots, n-1\}|=k-1$.
\end{lemma}
{\bf Proof. }
Let $x=(n-1)/k$. Then $0<x<n$, and $xk\equiv -1\ (\bmod\;n)$. Hence, $w=n-x$. Let us consider
\begin{eqnarray*}
\mathcal{B}&=&\{wk, (w+1)k, \cdots, (n-2)k, (n-1)k\}(\bmod\;n)\\
&=&\{-xk, -xk+k, \cdots, -xk+(x-2)k, -xk+(x-1)k\}(\bmod\;n)\\
&=&\{1, 1+k, \cdots, 1+(x-2)k, 1+(x-1)k\}\\
&=&\{n-xk, n-(x-1)k, \cdots, n-2k, n-k\}.
\end{eqnarray*}
Therefore, $\mathcal{B}\cap \{n-k, n-k+1, \cdots, n-1\}=\{n-k\}$. By Lemma \ref{lem_AB}, it follows that $$|\mathcal{A}\cap \{n-k, n-k+1, \cdots, n-1\}|=k-|\mathcal{B}\cap \{n-k, n-k+1, \cdots, n-1\}|=k-1.$$\done

\begin{theorem}\label{thm_k_n-1}
Let $g(x)$ be the dual function of the Coulter-Matthews bent function $Tr(ax^{d})$, where $a\in\mathbb{F}_{3^n}^{*}$.
If $k|n-1$ and $k>1$, then
$$g(x)=\sum_{j\in U_0}Tr\left((-1)^{N_2(j)+1}\eta(a)a^jx^{-jd}\right)+\sum_{j\in V_0}Tr\left((-1)^{N_2(j)}\eta(a)a^jx^{-jd}\right),$$
where $x^{-jd}=0$ if $x=0$, $U_0$ is defined by (\ref{eqn_U0}), and $V_0$ is defined by (\ref{eqn_V0}). Besides, there are $$\left(\left(\frac{1+\sqrt{5}}{2}\right)^{n-(n-1)/k+1}-\left(\frac{1-\sqrt{5}}{2}\right)^{n-(n-1)/k+1}\right)/\sqrt{5}$$ trace terms.
\end{theorem}
{\bf Proof. }
By Lemma \ref{lem_k_n-1}, $w=n-(n-1)/k$, and $|\mathcal{A}\cap \{n-k, n-k+1, \cdots, n-1\}|$ is even.
By Theorem \ref{thm_main_2}, the result follows.
\done

\begin{remark}
With notation as in Theorem \ref{thm_k_n-1}, the following two cases are especially interesting: 1) if $k=n-1$, then there are $\left(\left(\frac{1+\sqrt{5}}{2}\right)^n-\right.$ $\left.\left(\frac{1-\sqrt{5}}{2}\right)^n\right)/\sqrt{5}$ trace terms; 2) if $k=(n-1)/2$, then there are $\left(\left(\frac{1+\sqrt{5}}{2}\right)^{n-1}-\left(\frac{1-\sqrt{5}}{2}\right)^{n-1}\right)/\sqrt{5}$ trace terms.
\end{remark}

\begin{example}
Let $n=11$, and $k=5$. Then $k=(n-1)/2$, and $d=122$. By (\ref{eqn_U0}), $U_0=\{00001000010, $ $00001000210, 00001002010, 00001002210, 00001020010,
00001020210, 00001022010, 00001022210, 000012$\\ $00010, 00001200210, 00001202010, 00001202210, 00001220010, 00001220210, 00001222010, 00001222210, $\\ $00201000010, 00201020010, 00201200010, 00201220010, 02001000010, 02001000210, 02001200010, 020012$\\ $00210, 02201000010, 02201200010, 20001000010, 20001000210, 20001002010, 20001002210, 20201000010, $\\ $22001000010, 22001000210, 22201000010\}$, and by (\ref{eqn_V0}), $V_0=\{00001000012, 00001000212, 00001002012, $ $00001002212,
00001020012, 00001020212, 00001022012, 00001022212, 00201000012, 00201020012, 020010$\\ $00012, 02001000212, 02201000012, 20001000012, 20001000212, 20001002012, 20001002212, 20201000012, $\\ $22001000012, 22001000212, 22201000012\}$. By Theorem \ref{thm_k_n-1}, for any $a\in \mathbb{F}_{3^{11}}^{*}$, the dual function $g(x)$ of the Coulter-Matthews bent function $Tr(ax^{122})$ over $\mathbb{F}_{3^{11}}$ is given by
$$g(x)=\sum_{j\in U_0}Tr((-1)^{N_2(j)+1}\eta(a)a^jx^{-122j})+\sum_{j\in V_0}Tr((-1)^{N_2(j)}\eta(a)a^jx^{-122j}).$$
Totally, there are 55 trace terms.
\end{example}

\subsection{The Case of $(n-k)|n+1$ and $1<k<n-1$}\label{subsec_nk+1}

\begin{lemma}\label{lem_nk_n+1}
If $(n-k)|n+1$ and $1<k<n-1$, then $w=n-(n+1)/(n-k)$. Moreover, we have $|\mathcal{A}\cap \{n-k, n-k+1, \cdots, n-1\}|=k-\lfloor(k+1)/(n-k)\rfloor$.
\end{lemma}
{\bf Proof. }
Let $x=(n+1)/(n-k)$. Then $0<x<n$, and $xk\equiv -1\ (\bmod\;n)$. Hence, $w=n-x$. Let us consider
\begin{eqnarray*}
\mathcal{B}&=&\{wk, (w+1)k, (w+1)k, \cdots, (n-1)k\}(\bmod\;n)\\
&=&\{-xk, -xk+k, -xk+k, \cdots, -xk+(x-1)k\}(\bmod\;n)\\
&=&\{1, n+1-(-k), n+1-2(-k), \cdots, n+1-(x-1)(-k)\}(\bmod\;n)\\
&=&\{1, n+1-(n-k), n+1-2(n-k), \cdots, n+1-(x-1)(n-k)\}.
\end{eqnarray*}
Let $y=\lfloor(k+1)/(n-k)\rfloor$, where $\lfloor\cdot\rfloor$ is the floor function. Then, we get
$$\mathcal{B}\cap \{n-k, n-k+1, \cdots, n-1\}=\{n+1-(n-k), n+1-2(n-k), \cdots, n+1-y(n-k)\}.$$
By Lemma \ref{lem_AB}, it follows that
\begin{eqnarray*}
|\mathcal{A}\cap \{n-k, n-k+1, \cdots, n-1\}|&=&k-|\mathcal{B}\cap \{n-k, n-k+1, \cdots, n-1\}|\\
&=&k-\lfloor(k+1)/(n-k)\rfloor.
\end{eqnarray*}
\done

\begin{theorem}\label{thm_nk_n+1}
Let $g(x)$ be the dual function of the Coulter-Matthews bent function $Tr(ax^{d})$, where $a\in\mathbb{F}_{3^n}^{*}$.
Suppose that $(n-k)|n+1$ and $1<k<n-1$.

1) If $\lfloor(k+1)/(n-k)\rfloor$ is odd, then
$$g(x)=\sum_{j\in U_0}Tr\left((-1)^{N_2(j)+1}\eta(a)a^jx^{-jd}\right)+\sum_{j\in V_0}Tr\left((-1)^{N_2(j)}\eta(a)a^jx^{-jd}\right),$$
where $x^{-jd}=0$ if $x=0$, $U_0$ is defined by (\ref{eqn_U0}), and $V_0$ is defined by (\ref{eqn_V0}). Besides, there are $$\left(\left(\frac{1+\sqrt{5}}{2}\right)^{n-(n+1)/(n-k)+1}-\left(\frac{1-\sqrt{5}}{2}\right)^{n-(n+1)/(n-k)+1}\right)/\sqrt{5}$$ trace terms.

2) If $\lfloor(k+1)/(n-k)\rfloor$ is even, then
$$g(x)=\sum_{j\in U_1}Tr\left((-1)^{N_2(j)+1}\eta(a)a^jx^{-jd}\right)+\sum_{j\in V_1}Tr\left((-1)^{N_2(j)}\eta(a)a^jx^{-jd}\right),$$
where $x^{-jd}=0$ if $x=0$, $U_1$ is defined by (\ref{eqn_U1}), and $V_1$ is defined by (\ref{eqn_V1}). Besides, there are $$\left(\left(\frac{1+\sqrt{5}}{2}\right)^{(n+1)/(n-k)+1}-\left(\frac{1-\sqrt{5}}{2}\right)^{(n+1)/(n-k)+1}\right)/\sqrt{5}$$ trace terms.
\end{theorem}
{\bf Proof. }
By Lemma \ref{lem_nk_n+1}, $w=n-(n+1)/(n-k)$, and $|\mathcal{A}\cap \{n-k, n-k+1, \cdots, n-1\}|=k-\lfloor(k+1)/(n-k)\rfloor$.
By Theorem \ref{thm_main_2}, the result follows.
\done

\begin{corollary}\label{cor_nk_n+1}
With notation as in Theorem \ref{thm_nk_n+1}, let $n=mt+m-1$ and $k=(m-1)t+m-2$, where $m\geq 3$ and $t\geq 1$.

1) If $m$ is even and $t$ is odd, then
$$g(x)=\sum_{j\in U_0}Tr\left((-1)^{N_2(j)+1}\eta(a)a^jx^{-jd}\right)+\sum_{j\in V_0}Tr\left((-1)^{N_2(j)}\eta(a)a^jx^{-jd}\right),$$
where $x^{-jd}=0$ if $x=0$, and there are $\left(\left(\frac{1+\sqrt{5}}{2}\right)^{n-m+1}-\left(\frac{1-\sqrt{5}}{2}\right)^{n-m+1}\right)/\sqrt{5}$ trace terms.

2) If $m$ is odd, then
$$g(x)=\sum_{j\in U_1}Tr\left((-1)^{N_2(j)+1}\eta(a)a^jx^{-jd}\right)+\sum_{j\in V_1}Tr\left((-1)^{N_2(j)}\eta(a)a^jx^{-jd}\right),$$
where $x^{-jd}=0$ if $x=0$, and there are $\left(\left(\frac{1+\sqrt{5}}{2}\right)^{m+1}-\left(\frac{1-\sqrt{5}}{2}\right)^{m+1}\right)/\sqrt{5}$ trace terms.
\end{corollary}
{\bf Proof. }
One may check that $\gcd(n, k)=1$, and $(n-k)|n+1$. Moreover, $k$ is odd if and only if $m$ is odd, or $m$ is even and $t$ is odd. In addition, $\lfloor(k+1)/(n-k)\rfloor=m-1$. By Theorem \ref{thm_nk_n+1}, the result follows. \done

\begin{remark}
With notation as in Corollary \ref{cor_nk_n+1}, the following three cases are especially interesting. 1) If $m=3$, then there are $\left(\left(\frac{1+\sqrt{5}}{2}\right)^{4}-\left(\frac{1-\sqrt{5}}{2}\right)^{4}\right)/\sqrt{5}=3$ trace terms. Actually, this is just the case of Theorem \ref{thm_3_term_1}. 2) If $m=5$, then there are $\left(\left(\frac{1+\sqrt{5}}{2}\right)^{6}-\left(\frac{1-\sqrt{5}}{2}\right)^{6}\right)/\sqrt{5}=8$ trace terms. Hence, we find a new class of bent functions with 8 terms. 3) If $m=7$, then there are $\left(\left(\frac{1+\sqrt{5}}{2}\right)^{8}-\left(\frac{1-\sqrt{5}}{2}\right)^{8}\right)/\sqrt{5}=21$ trace terms. Hence, we find a new class of bent functions with 21 terms.
\end{remark}

\subsection{The Case of $(n-k)|n-1$ and $1<k<n-1$}\label{subsec_nk-1}

\begin{lemma}\label{lem_nk_n-1}
If $(n-k)|n-1$ and $1<k<n-1$, then $w=(n-1)/(n-k)$. Moreover, we have $|\mathcal{A}\cap \{n-k, n-k+1, \cdots, n-1\}|=\lfloor k/(n-k)\rfloor$.
\end{lemma}
{\bf Proof. }
Let $x=(n-1)/(n-k)$. Then $0<x<n$, and $xk\equiv 1\ (\bmod\;n)$. Hence, $w=x$. As a consequence,
\begin{eqnarray*}
\mathcal{A}&=&\{0, k, 2k, \cdots, (w-1)k\}(\bmod\;n)\\
&=&\{0, -(n-k), -2(n-k), \cdots, -(w-1)(n-k)\}(\bmod\;n)\\
&=&\{0, n-(n-k), n-2(n-k), \cdots, n-(w-1)(n-k)\}.
\end{eqnarray*}
Let $y=\lfloor k/(n-k)\rfloor$. Then $$\mathcal{A}\cap \{n-k, n-k+1, \cdots, n-1\}=\{n-(n-k), n-2(n-k), \cdots, n-y(n-k)\}.$$
Therefore, we get $|\mathcal{A}\cap \{n-k, n-k+1, \cdots, n-1\}|=\lfloor k/(n-k)\rfloor$.\done

\begin{theorem}\label{thm_nk_n-1}
Let $g(x)$ be the dual function of the Coulter-Matthews bent function $Tr(ax^{d})$, where $a\in\mathbb{F}_{3^n}^{*}$.
Suppose that $(n-k)|n-1$ and $1<k<n-1$.

1) If $\lfloor k/(n-k)\rfloor$ is even, then
$$g(x)=\sum_{j\in U_0}Tr\left((-1)^{N_2(j)+1}\eta(a)a^jx^{-jd}\right)+\sum_{j\in V_0}Tr\left((-1)^{N_2(j)}\eta(a)a^jx^{-jd}\right),$$
where $x^{-jd}=0$ if $x=0$, $U_0$ is defined by (\ref{eqn_U0}), and $V_0$ is defined by (\ref{eqn_V0}). Besides, there are $$\left(\left(\frac{1+\sqrt{5}}{2}\right)^{(n-1)/(n-k)+1}-\left(\frac{1-\sqrt{5}}{2}\right)^{(n-1)/(n-k)+1}\right)/\sqrt{5}$$ trace terms.

2) If $\lfloor k/(n-k)\rfloor$ is odd, then
$$g(x)=\sum_{j\in U_1}Tr\left((-1)^{N_2(j)+1}\eta(a)a^jx^{-jd}\right)+\sum_{j\in V_1}Tr\left((-1)^{N_2(j)}\eta(a)a^jx^{-jd}\right),$$
where $x^{-jd}=0$ if $x=0$, $U_1$ is defined by (\ref{eqn_U1}), and $V_1$ is defined by (\ref{eqn_V1}). Besides, there are $$\left(\left(\frac{1+\sqrt{5}}{2}\right)^{n-(n-1)/(n-k)+1}-\left(\frac{1-\sqrt{5}}{2}\right)^{n-(n-1)/(n-k)+1}\right)/\sqrt{5}$$ trace terms.
\end{theorem}
{\bf Proof. }
By Lemma \ref{lem_nk_n-1}, $w=(n-1)/(n-k)$, and $|\mathcal{A}\cap \{n-k, n-k+1, \cdots, n-1\}|=\lfloor k/(n-k)\rfloor$.
By Theorem \ref{thm_main_2}, the result follows.
\done

\begin{corollary}\label{cor_nk_n-1}
With notation as in Theorem \ref{thm_nk_n-1}, let $n=mt+1$ and $k=(m-1)t+1$, where $m\geq 3$ and $t\geq 1$.

1) If $m$ is odd, then
$$g(x)=\sum_{j\in U_0}Tr\left((-1)^{N_2(j)+1}\eta(a)a^jx^{-jd}\right)+\sum_{j\in V_0}Tr\left((-1)^{N_2(j)}\eta(a)a^jx^{-jd}\right),$$
where $x^{-jd}=0$ if $x=0$, and there are $\left(\left(\frac{1+\sqrt{5}}{2}\right)^{m+1}-\left(\frac{1-\sqrt{5}}{2}\right)^{m+1}\right)/\sqrt{5}$ trace terms.

2) If $m$ is even and $t$ is even, then
$$g(x)=\sum_{j\in U_1}Tr\left((-1)^{N_2(j)+1}\eta(a)a^jx^{-jd}\right)+\sum_{j\in V_1}Tr\left((-1)^{N_2(j)}\eta(a)a^jx^{-jd}\right),$$
where $x^{-jd}=0$ if $x=0$, and there are $\left(\left(\frac{1+\sqrt{5}}{2}\right)^{n-m+1}-\left(\frac{1-\sqrt{5}}{2}\right)^{n-m+1}\right)/\sqrt{5}$ trace terms.

\end{corollary}
{\bf Proof. }One may check that $\gcd(n, k)=1$, and $(n-k)|n-1$. Moreover, $k$ is odd if and only if $m$ is odd, or $m$ is even and $t$ is even. In addition, $\lfloor k/(n-k)\rfloor=m-1$. By Theorem \ref{thm_nk_n+1}, the result follows.\done

\begin{remark}
With notation as in Corollary \ref{cor_nk_n-1}, similar to the case of Corollary \ref{cor_nk_n+1}, the following three cases are especially interesting. 1) If $m=3$, then there are 3 trace terms. Actually, this is just the case of Theorem \ref{thm_3_term_2}. 2) If $m=5$, then there are 8 trace terms.
Hence, compared with Corollary \ref{cor_nk_n+1}, we find another class of bent functions with 8 terms.
3) If $m=7$, then there are 21 trace terms. Hence, compared with Corollary \ref{cor_nk_n+1}, we find another class of bent functions with 21 terms.
\end{remark}

\section{Concluding Remarks}\label{sec_con}

In this paper, using new combinatorial technique, we find an explicit expression for the dual function of the Coulter-Matthews bent functions
for all cases of $k$ and $n$. For four special cases, this expression can be improved further. As a consequence, we find many classes of ternary bent functions which have not been reported in the literature previously. In particular, new classes of ternary bent functions with only 8 or 21 trace terms
have been dug out. In case of the correlation distribution of the Coulter-Matthews decimation \cite{NHK06}, the new findings of this paper can be applied directly. One left problem is to determine the value of $|\mathcal{A}\cap \{n-k, n-k+1, \cdots, n-1\}|$
for any $n$ and $k$ with $k$ odd and $\gcd(n, k)=1$.



\begin{thebibliography}{100}

\bibitem{CaChDo00}
A. Canteaut, P. Charpin, and H. Dobbertin, Binary $m$-sequences with three-valued crosscorrelation:
A proof of Welch's conjecture, \emph{{IEEE} Trans. Inf. Theory}, vol. 46, no. 1, pp. 4-8, Jan. 2000.

\bibitem{CM97}
R. S. Coulter and R.W. Matthews, Planar functions and planes of Lenz-Barlotti class II, {\em Des., Codes Cryptogr.},
vol. 10, no. 2, pp. 167-184, Feb. 1997.

\bibitem{CM97-1}
R. S. Coulter and R. W. Matthews, Bent polynomials over finite fields, {\em Bulletin of the Australian Mathematical Society}, vol. 56, pp. 429-437, 1997.

\bibitem{Dillon72}
J. Dillon, A survey of bent functions, In {\em NSA Technical Journal Special Issue}, pp. 191-215, 1972.

\bibitem{FL07}
K. Feng and J. Luo, Value distributions of exponential sums from perfect nonlinear functions and their applications,
{\em IEEE Trans. Inf. Theory}, vol. 53, no. 9, pp. 3035-3041, Sep. 2007.

\bibitem{GG05}
S. W. Golomb and G. Gong, {\em Signal Design for Good Correlation: For Wireless Communication, Cryptography and Radar.} Cambridge,
U.K.: Cambridge Univ. Press, 2005.

\bibitem{GHHK12}
G. Gong, T. Helleseth, H. Hu, and A. Kholosha, On the dual of certain ternary weakly regular bent functions,
{\em IEEE Trans. Inf. Theory}, vol. 58, no. 4, pp. 2237-2243, Apr. 2012.

\bibitem{HeHoKhWaXi09}
T. Helleseth, H. D. L. Hollmann, A. Kholosha, Z. Wang, and Q. Xiang, Proofs
  of two conjectures on ternary weakly regular bent functions, {\em IEEE
  Trans. Inf. Theory}, vol. 55, no. 11, pp. 5272-5283, Nov. 2009.

\bibitem{HeKh06_1}
T. Helleseth and A. Kholosha, Monomial and quadratic bent functions over the
  finite fields of odd characteristic, {\em IEEE Trans. Inf. Theory},
  vol. 52, no. 5, pp. 2018-2032, May 2006.

\bibitem{HeKh07}
T. Helleseth and A. Kholosha, On the dual of monomial quadratic $p$-ary bent functions, in
  {\em Sequences, Subsequences, and Consequences}, ser. Lecture Notes in
  Computer Science, S. Golomb, G. Gong, T. Helleseth, and H. Y. Song, Eds.,
  vol. 4893. Berlin: Springer-Verlag, pp. 50-61, 2007.

\bibitem{HeKh10}
T. Helleseth and A. Kholosha, New binomial bent functions over the finite
  fields of odd characteristic, {\em IEEE Trans. Inf. Theory}, vol. 56,
  no. 9, pp. 4646-4652, Sep. 2010.

\bibitem{HK98}
T. Helleseth and P. V. Kumar, Sequences with low correlation, in
{\em Handbook of Coding Theory}, V. S. Pless and W. C. Huffman, Eds.
Amsterdam, The Netherlands: Elsevier Science, 1998, pp. 1765-1853.

\bibitem{HoXi01}
H. D. L. Hollmann and Q. Xiang, A proof of the Welch and Niho conjectures on
cross-correlations of binary $m$-sequences, {\em Finite Fields Appl.}, vol. 7, no. 2, pp. 253-286, Apr. 2001.

\bibitem{Ho08}
X. D. Hou, On the dual of a {C}oulter-{M}atthews bent function, {\em Finite Fields Appl.}, vol. 14, no. 2, pp. 505-514, Apr. 2008.

\bibitem{HSGH14}
H. Hu, S. Shao, G. Gong, and T. Helleseth, The proof of Lin's conjecture via the decimation-Hadamard transform,
{\em IEEE Trans. on Inform. Theory}, vol. 60, no. 8, pp. 5054-5064, Aug. 2014.

\bibitem{HZS17}
H. Hu, Q. Zhang, and S. Shao, On the dual of the Coulter-Matthews bent functions, {\em IEEE Trans. on Inform. Theory},
vol. 63, no. 4, pp. 2454-2463, Apr. 2017.

\bibitem{KJNH03}
Y. S. Kim, J. W. Jang, J. S. No, and T. Helleseth, On $p$-ary bent functions
defined on finite fields, in {\em Mathematical Properties of Sequences
and other Combinatorial Structures}. ser. The Kluwer International Series
in Engineering and Computer Science, J. S. No, H. Y. Song, T.
Helleseth, and P. V. Kumar, Eds. Dordrecht, The netherlands: Kluwer
Academic Publishers, pp. 65-76, 2003.

\bibitem{KM91}
P. V. Kumar and O. Moreno, Prime-phase sequences with periodic correlation properties better than binary sequences, {\em IEEE Trans. Inf.
Theory}, vol. 37, no. 3, pp. 603-616, May 1991.

\bibitem{KuScWe85}
P. V. Kumar, R. A. Scholtz, and L. R. Welch, Generalized bent functions and their properties,
{\em J. Combin. Theory. Ser. A}, vol. 40, no. 1, pp. 90-107, Sep. 1985.

\bibitem{LN83}
R. Lidl and H. Niederreiter, {\em Finite Fields}. Reading, MA: Addison-Wesley, 1983, now distributed by Cambridge Univ. Press.

\bibitem{LK92}
S. C. Liu and J. J. Komo, Nonbinary Kasami sequences over GF($p$), {\em IEEE Trans. Inf. Theory}, vol. 38, no. 4, pp. 1409-1412, Jul. 1992.

\bibitem{Mc72}
R. J. McEliece, Weight congruences for $p$-ary cyclic codes, \emph{Discrete  Math.}, vol. 3, no. 1-3, pp. 177-192, 1972.

\bibitem{Mesnager16}
S. Mesnager, {\em Bent Functions: Fundamentals and Results}, Springer, Switzerland, 2016.

\bibitem{NHK06}
G. Ness, T. Helleseth, and A. Kholosha, On the correlation distribution of the Coulter-Matthews decimation,
{\em IEEE Trans. Inf. Theory}, vol. 52, no. 5, pp. 2241-2247, May 2006.

\bibitem{Rothaus76}
O. S. Rothaus, On ``bent" functions, {\em J. Combin. Theory, Ser. A}, vol. 20, no. 3, pp. 300-305, May 1976.

\end{thebibliography}
\end{document}